\documentclass{appolb}
\usepackage{amsmath,amsfonts,amssymb}
\usepackage{relsize}
\usepackage[dvips]{graphicx}

\begin{document}
\title{The large-$N_c$ masses of light scalar mesons from QCD sum rules for linear radial spectrum\thanks{Presented at Excited QCD 2018.}}
\author{S. S. Afonin and T. D. Solomko\address{Saint Petersburg State University, 7/9 Universitetskaya nab., St.Petersburg, 199034, Russia}}
\maketitle

\begin{abstract}
We discuss a calculation of large-$N_c$ masses of light scalar mesons from the QCD sum rules.
Two methods based on the use of linear radial Regge trajectories are presented.
We put a special emphasis on the appearance of pole near 0.5~GeV in the scalar isoscalar channel
which emerges in both methods and presumably corresponds to the scalar sigma-meson.
\end{abstract}


\section{Introduction}

It is widely known that the physics of non-perturbative strong
interactions is enciphered in the values of hadron masses. This intricate physics
is especially pronounced in the hadrons consisting of
$u$- and $d$-quarks as their masses $m_{u,d}$ are much less than
the non-perturbative scale $\Lambda_{\text{QCD}}$. At the same time,
the given hadrons shape the surrounding world. Aside from the
nucleons and pions, an important role is played by the scalar
sigma-meson which is responsible for the main part of the
nucleon attraction potential. In the particle physics, the given
resonance is identified as $f_0(500)$~\cite{pdg} and is indispensable
for description of the chiral symmetry breaking in many
phenomenological field models describing the strong interactions.
The scalar sector below and near 1~GeV is perhaps the most  difficult for traditional
approaches in the hadron spectroscopy. The usual quark model faces serious problems
in explaining the existence and properties of light scalar mesons, perhaps due to a strong
admixture of glueball component. Despite the recent
progress in description of these states by dispersive methods~\cite{pdg,pelaez}
the scalar sector still remains puzzling.

The physical characteristics of hadrons are encoded in various
correlation functions of corresponding hadron currents. Perhaps
the most important characteristics is the hadron mass. The
calculation of a hadron mass from first principles consists in
finding the relevant pole of two-point correlator $\left\langle
JJ\right\rangle$, where the current $J$ is built from the quark
and gluon fields and interpolates the given hadron. For instance,
if the scalar isoscalar state $f_0$ represents an ordinary light
non-strange quark-antiquark meson, its current should be
interpolated by the quark bilinear $J=\bar{q}q$, where $q$ stays
for the $u$ or $d$ quark. In the real QCD, the straightforward
calculations of correlators are possible only in the framework of
lattice simulations which are still rather restricted.

It is usually believed that confinement in QCD leads to approximately linear
radial Regge trajectories (see, e.g.,~\cite{phen}). The most important
quantity in this picture is the slope of trajectories. The slope is expected
to be nearly universal as arising from flavor-independent non-perturbative
gluodynamics which thereby sets a mass scale for light hadrons.

Among the phenomenological approaches to the hadron spectroscopy, the method
of spectral sum rules~\cite{svz} is perhaps the most related with QCD.
In many cases, it permits to calculate reliably the masses of ground
states on the radial trajectories. This method exploits some information from QCD via
the Operator Product Expansion (OPE) of correlation
functions~\cite{svz}. On the other hand, one assumes a certain
spectral representation for a correlator in question. Typically
the representation is given by the ansatz "one infinitely narrow
resonance + perturbative continuum". Such an approximation is very
rough but works well phenomenologically in many
cases. Theoretically the zero-width approximation arises in the large-$N_c$
limit of QCD~\cite{hoof}. In this limit, the
only singularities of the two-point correlation function of a
hadron current $J$ are one-hadron states. For instance,
the two-point correlator for $J=\bar{q}q$ has the following form to
lowest order in $1/N_c$ (in the momentum space),
\begin{equation}
\label{20}
\Pi_S(q^2)=\left\langle J^S(q)J^S(-q)\right\rangle=\sum_n\frac{G_n^2M_S^2(n)}{q^2-M_S^2(n)},
\end{equation}
where the residues appear from the definition of the matrix element $\langle0|J^S|n\rangle=G_nM_S(n)$.
The OPE of the correlator~\eqref{20} in the large-$N_c$
limit and to the lowest order in the perturbation theory reads~\cite{rry}
\begin{equation}
\label{27}
\Pi_S(Q^2)=\frac{3Q^2}{16\pi^2}\log{\frac{Q^2}{\mu^2}}+ \frac{3}{2Q^2}m_q\langle\bar{q}q\rangle
-\frac{\alpha_s}{16\pi}\frac{\langle G^2\rangle}{Q^2}
-\frac{11}{3}\pi\alpha_s\frac{\langle\bar{q}q\rangle^2}{Q^4}+\dots,
\end{equation}
where $\langle G^2\rangle$ and $\langle\bar{q}q\rangle$ denote the
gluon and quark vacuum condensate, respectively. According to the
main assumption of classical QCD sum rules~\cite{svz}, these vacuum
characteristics are universal, i.e., their values do not depend on
the quantum numbers of a hadron current $J$ (the method is not
applicable otherwise).

In the present talk we will demonstrate how all these ideas can be used
for calculation of large-$N_c$ masses of light scalar mesons.

\section{Scalar sum rules: Some results}

We will assume the linear radial spectrum with universal slope
\begin{equation}
\label{21}
M_S^2(n)=\Lambda^2(n+m_s^2), \qquad n=0,1,2,\dots,
\end{equation}
and (for consistency with the OPE): $G_n=G$.
With the linear ansatz~\eqref{21} for
the radial mass spectrum, the expression~\eqref{20} can be summed analytically,
expanded at large $Q^2=-q^2$ and compared with the corresponding
OPE in QCD. Thus one obtains a set of sum rules. Similar large-$N_c$ sum rules were considered many
times in the past for vector, axial, scalar and pseudoscalar channels (see, e.g., Refs. in~\cite{sr}).

As {\it apriori} we do not know reliably the radial Regge behavior
of scalar masses, two simple possibilities can be considered: (I)
The ground $n=0$ state lies on the linear trajectory~\eqref{21};
(II) The state $n=0$, below called $\sigma$, is not described by
the linear spectrum~\eqref{21}. The second assumption looks more
physical. Within the latter assumption, the mass of $\sigma$-meson
can be derived as a function of the intercept parameter $m_s^2$
(we refer to Ref.~\cite{we1} for details, the chiral limit is considered),
\begin{equation}
\label{33}
M_\sigma^2=\frac{\frac{1}{16\pi^2}\Lambda^6m_s^2\left(m_s^2+\frac12\right)\left(m_s^2+1\right)+
\frac{11}{3}\pi\alpha_s\langle\bar{q}q\rangle^2}
{\frac{3}{32\pi^2}\Lambda^4\left(m_s^4+m_s^2+\frac16\right)+\frac{\alpha_s}{16\pi}\langle G^2\rangle}.
\end{equation}
Substituting the physical values of vacuum condensates and numerical value for slope $\Lambda^2$ obtained
from a solution of QCD sum rules, the mass function~\eqref{33} is displayed in Fig.~1~\cite{we1}.
\begin{figure}[ht]
\center{\includegraphics[width=0.7\linewidth]{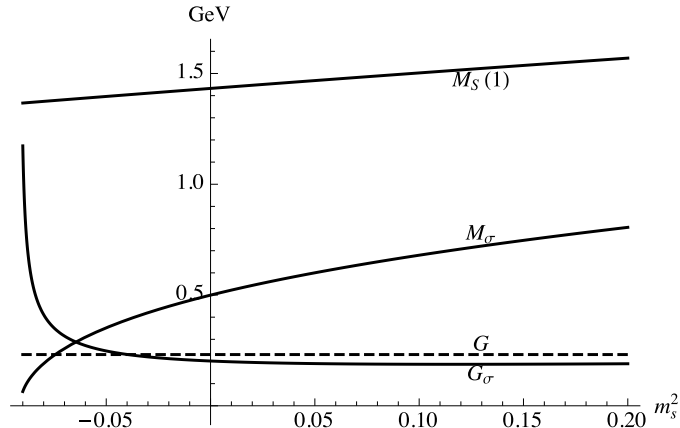}}
\vspace{-0.3cm}
\caption{The values of $M_\sigma$, $G_\sigma$, $G$, and
the first state on the scalar trajectory $M_S(1)$ as a
function of dimensionless intercept $m_s^2$.}
\end{figure}

The mass of the first radially excited state $M_S(1)$ is rather stable and
seems to reproduce the mass of $a_0(1450)$-meson,
$M_{a_0(1450)}=1474\pm19$~MeV~\cite{pdg}. Its isosinglet partner
(the candidates is $f_0(1370)$) should be degenerate with
$a_0(1450)$ in the planar limit.

The plot in Fig.~1 demonstrates that the actual prediction for
$M_\sigma$ is rather sensitive to the intercept of scalar linear
trajectory, though initially $M_\sigma$ is not described by the
linear spectrum~\eqref{21}. And {\it vice versa}, the expected
value of $M_\sigma$ (around $0.5$~GeV~\cite{pdg}) imposes a strong
bound on the allowed values of intercept $m_s^2$. The plot in
Fig.~1 shows that $m_s^2$ is likely close to zero.

Thus, interpolating the scalar states by the simplest quark bilinear current,
we predict a light scalar resonance with mass about $500\pm100$~MeV
which is a reasonable candidate for the scalar sigma-meson $f_0(500)$~\cite{pdg}.

\section{Borelized scalar sum rules: Some results}

The original QCD sum rules made use of the Borel transformation~\cite{svz},
\begin{equation}
\label{6}
L_M\Pi(Q^2)=\lim_{\substack{Q^2,n\rightarrow\infty\\Q^2/n=M^2}}\frac{1}{(n-1)!}(Q^2)^n\left(-\frac{d}{dQ^2}\right)^n\Pi(Q^2),
\end{equation}
The borelized version has a number of advantages and can be applied to
our large-$N_c$ case. The details are contained in Ref.~\cite{we2}.
In short, the mass of ground scalar meson $m_0\equiv M_S(0)$ as a function of Borel parameter
is shown in Fig.~2. It is seen that there are two solutions with "Borel window" extending to infinity.
\begin{figure}[ht]
    \includegraphics[scale=0.7]{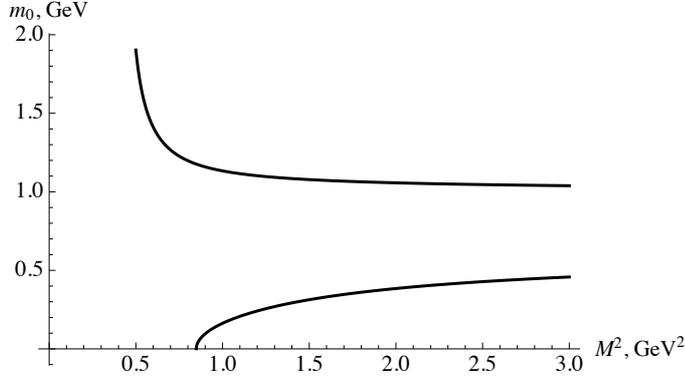}
    \caption{\small The mass of ground scalar meson $m_0\equiv M_S(0)$ as a function of Borel parameter at $\Lambda^2=1.38\,\text{GeV}^2$~\cite{we2}.}
\end{figure}
The corresponding asymptotic values are given by
\begin{equation}
\label{11}
M_S^2(0)=\frac{\Lambda^2}{2}\pm\frac{1}{2}\sqrt{\frac{\Lambda^4}{3}-64\pi^2\left(m_q\langle\bar{q}q\rangle + \frac{\alpha_s}{24\pi}\langle G^2\rangle\right)}.
\end{equation}

The heavier state corresponds to the ground scalar mass in the standard QCD sum rules.
Normalizing this mass to the value $m_{f_0}=1.00\pm0.03$~GeV extracted from these canonical sum rules~\cite{rry}
we predict the value of slope for the scalar trajectory, $\Lambda^2_{f_0}=1.38\pm0.07$~GeV$^2$, which is
used in Fig.~2. We obtain then the mass of the lightest scalar state, $M_\sigma\approx0.62$~GeV.

We arrive thus at the conclusion that our method predicts two
parallel scalar trajectories. The ground state on the first trajectory
can be identified with $f_0(980)$ and on the second one with $f_0(500)$~\cite{pdg}.
The existence of two parallel radial scalar trajectories seems to agree
with the experimental data~\cite{phen}. The masses of predicted radial
states and a tentative comparison with the observed scalar mesons for two
trajectories are displayed in Tables~1 and~2, correspondingly.

\begin{table}[ht]
\caption{\small The radial spectrum of the first $f_0$-trajectory for the slope $\Lambda^2=1.38\pm0.07\,\text{GeV}^2$.
The first 5 predicted states are tentatively assigned to the resonances $f_0(980)$, $f_0(1500)$,
$f_0(2020)$, $f_0(2200)$, and $X(2540)$~\cite{pdg}.}
\begin{center}
$\begin{array}{|c|c|c|c|c|c|}
 \hline
 n & 0 & 1 & 2 & 3 & 4\\
 \hline
 m_{f_0}\,\text{(th 1)} & 1000\pm30 & 1540\pm20 & 1940\pm40 & 2270\pm50 & 2560\pm50 \\
 \hline
 m_{f_0}\,\text{(exp 1)} & 990 \pm 20 & 1504 \pm 6 & 1992 \pm 16 & 2189 \pm 13 & 2539 \pm 14^{+38}_{-14} \\
 \hline
\end{array}$
\end{center}
\end{table}

\begin{table}[ht]
\caption{\small The radial spectrum of the second $f_0$-trajectory for the slope $\Lambda^2=1.38\pm0.07\,\text{GeV}^2$.
The first 5 predicted states are tentatively assigned to the resonances $f_0(500)$, $f_0(1370)$,
$f_0(1710)$, $f_0(2100)$, and $f_0(2330)$~\cite{pdg}.}
\begin{center}
$\begin{array}{|c|c|c|c|c|c|}
 \hline
 n & 0 & 1 & 2 & 3 & 4\\
 \hline
 m_{f_0}\,\text{(th 2)} & 620 & 1330\pm30 & 1780\pm40 & 2130\pm50 & 2430\pm60 \\
 \hline
 m_{f_0}\,\text{(exp 2)} & 400\text{--}550 & 1200\text{--}1500 & 1723^{+6}_{-5} & 2101 \pm 7 & 2300\text{--}2350 \\
 \hline
\end{array}$
\end{center}
\end{table}

\section*{Acknowledgements}

The present work was supported by a Saint-Petersburg State University
travel grant and by the RFBR grant 16-02-00348-a.

\end{document}